\documentclass{elsart}
\usepackage{epsfig,feynmp}
\def\be{\begin{equation}}
\def\ee{\end{equation}}
\def\beq{\begin{eqnarray}} 
\def\eeq{\end{eqnarray}}

\newcommand{\gsim}{\raisebox{-4pt}{$\,\stackrel{\textstyle
                             >}{\sim}\,$}} 
\renewcommand{\d}{{\rm d}}
\begin{document}
\begin{frontmatter}

\begin{flushright}
                      WUB 97-28 
%                     \\ hep-ph/9711231 
\end{flushright}
\vspace*{2cm}

\title{Flavor Symmetry Breaking and Mixing Effects \\
       in the $\eta \gamma$ and $\eta'\gamma$ Transition Form Factors}

\author{Thorsten Feldmann} and \author{Peter Kroll}

\address{Department of Theoretical Physics, University of Wuppertal,\\
          D-42097 Wuppertal, Germany}

\journal{?}
\thanks {e-Mail: {\tt kroll@theorie.physik.uni-wuppertal.de}}
\vspace*{2cm}

\begin{abstract}
{\bf Abstract}. We reanalyze the $\eta\gamma$ and $\eta'\gamma$  
transition form factors within the modified hard scattering approach 
on the basis of new experimental data from CLEO \cite{CLEO97} and L3 
\cite{L397}.  Our approach perfectly describes the experimental data 
over a wide range of the virtuality of the probing photon, 
$1~$GeV$^2\leq Q^2 \leq 15$~GeV$^2$. The analysis provides hints that 
the conventional flavor octet-singlet scheme for the $\eta$-$\eta'$ 
mixing is too simple.  A more general mixing scheme on the other hand,
involving two mixing angles, leads to a very good description of the
transition form factors and also accounts for the two-photon decay
widths of the $\eta$ and $\eta'$ mesons as well as for the ratio of the
widths for the $J/\psi \to \eta'\gamma$ and $J/\psi \to \eta\gamma$ 
decays. We also investigate the questions of possible deviations of 
the $\eta$ and $\eta'$ distribution amplitudes from the asymptotic 
form and of eventual intrinsic charm in the $\eta$ and $\eta'$ mesons.
We estimate the charm decay constant of the $\eta'$ meson
to lie within the range $-65$~MeV${}\leq f_{\eta'}^c\leq 15$~MeV.
\end{abstract}

\end{frontmatter}
\newpage
\baselineskip1.5em
\pagestyle{plain}
%%%%%%%%%%%%%%%%%%%%%%%%%%%%%%%%%%%%%%%%%%%%%%%%%%%%%%%%%%%%%%%%%%%
\setcounter{page}{1} 
%\vspace*{-0.6cm}
\section{Introduction}
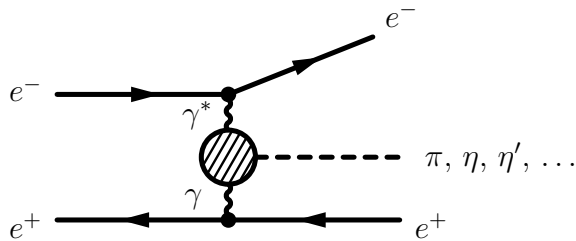
\begin{figure}[hbt]
\begin{center}
\vspace{2em}
\begin{fmfgraph*}(6,4)
\fmfpen{thick}
\fmfleft{bb,ep1,vv,em1,cc} \fmfright{aa,ep2,etac,xx,em2}
\fmf{fermion}{ep2,v2,ep1}
\fmf{fermion}{em1,v1}
\fmf{phantom}{v1,xx}
\fmffreeze
\fmf{fermion}{v1,em2}
\fmf{photon,label=$\gamma^*$}{v1,v}
\fmf{photon,label=$\gamma$  }{v,v2}
\fmffreeze
\fmf{dashes}{v,etac}
\fmfdot{v1,v2}
\fmfvn{label=$e^+$}{ep}{2} 
\fmfvn{label=$e^-$}{em}{2}
\fmfv{label=$\pi$,, $\eta$,, $\eta'$,, \noexpand\ldots}{etac} 
\fmfv{dec.sh=circle,dec.filled=.3,dec.si=.15w}{v}      
\end{fmfgraph*}
\end{center}
\caption{Meson-photon transition form factors in $e^+e^-$ collisions.}
\label{Profig}
\end{figure}

\vspace*{-0.2cm}
%%%%%%%%%%%%%%%%%%%%%%%%%%%%%%%%%%%%%%%%%%%%%%%%%%%%%%%%%%%%%%%%%%%
\begin{fmffile}{etagammaffpic}
\unitlength0.8cm
In 1995 the CLEO collaboration has presented their preliminary data on
pseudoscalar meson-photon transition form factors  (see
Fig.\ \ref{Profig}) at large momentum transfer, $Q^2$, for the first time
\cite{CLEO95}. Since then these form factors attracted the interest of
many theoreticians, and it can be said now that the CLEO measurement
has strongly stimulated the field of hard exclusive reactions. One of
the exciting aspects of the $\pi\gamma$ form factor is that it
possesses a well-established asymptotic behavior \cite{WaZe73,BrLe80},
namely $F_{\pi\gamma} \to \sqrt2f_\pi/Q^2$ where $f_\pi(=131 {\rm
MeV})$ is the decay constant of the pion.  At the upper end of the
measured momentum transfer range  the CLEO data \cite{CLEO95,CLEO97} 
only deviate by about 15\% from that limiting value. Many theoretical 
papers are devoted to the explanation of that little difference.  The 
perhaps most important outcome of these analyses, as far as they are 
based upon perturbative approaches (see e.g.\ \cite{KrRa96,JaKrRa96,MuRa97}), 
is the rather precise determination of the pion's light-cone wave
function.  It turns out that the pion's distribution amplitude, i.e.\
its wave function integrated over transverse momentum, is close to the
asymptotic form.  This result has far-reaching consequences for the
explanation of many hard exclusive reactions in which pions
participate (see, for instance, \cite{BrJiPaRo97,Bolz:1997}).

The situation is much more complicated for the $\eta\gamma$ and 
$\eta'\gamma$ transition form factors than in the $\pi\gamma$ case. In
general there are at least four independent wave functions associated
with the $\eta$ and $\eta'$ valence Fock states, since one has to
consider $SU(3)_F$-singlet (octet) admixtures to the $\eta(\eta')$
mesons.  Moreover, on account of the $U(1)_A$ anomaly, there is also
the possibility of intrinsic charm and gluon admixtures.
Correspondingly numerous are the decay constants being related to the
configuration space wave functions at the origin.  A full-fledged
analysis of the transition form factors, taking into account all these
components of the $\eta$ and $\eta'$ mesons, is beyond feasibility. 
Although the recent large momentum transfer data on the $\eta\gamma$ and
$\eta'\gamma$ transition form factors measured by  CLEO \cite{CLEO97}
and L3 \cite{L397} allow a more refined analysis of these processes
than it was possible previously \cite{JaKrRa96,Anisovich97},
additional phenomenological constraints as well as simplifying
assumptions on the wave functions and the decay constants
are still required. Thus, for instance, we will use the two-photon 
decays of  the $\eta$ and $\eta'$ mesons as a constraint . The chiral 
anomaly combined with the PCAC hypothesis relates the decay constants 
to the decay widths for these two processes. Another constraint is 
offered by the ratio $R_{J/\psi}$ of the $J/\psi \to \eta'\gamma$ and 
$J/\psi \to \eta\gamma$ decay widths, since it can also be expressed 
in terms of the decay constants.

An obvious possibility to reduce the degrees of freedom in the
analysis is the use of the conventional $SU(3)_F$ octet-singlet mixing
scheme. Leaving aside eventual  charm and gluon components, only two
independent wave functions remain thereby and, hence, only two decay
constants.  The relative strength of the singlet and octet components
in the physical mesons is then controlled by the pseudoscalar mixing
angle $\theta_P$. This mixing scheme has been used in previous
analyses of the $\eta\gamma$, $\eta'\gamma$ transition form factors
throughout \cite{JaKrRa96,Anisovich97,Bijnensetal92}.
As it will turn out from our analysis of the new large momentum
transfer form factor data \cite{CLEO97,L397}, that mixing scheme is 
inadequate; it leads to  inconsistencies with results from chiral 
perturbation theory \cite{GaLe85,DoHoLi86} and is in conflict with 
$R_{J/\psi}$.  An ansatz, however, where the four relevant wave 
functions are assumed to have the same form but different values at the
origin, i.e.\ different decay constants, meets all requirements: It
leads to  very good results for the transition form factors, the decay
widths for $\eta(\eta')\to \gamma\gamma$ and for $R_{J/\psi}$. The
decay constants determined in this analysis are in good agreement with
the recent results from chiral perturbation theory in which also the
conventional octet-singlet mixing scheme is given up
\cite{Leutwyler97}.

Another interesting problem that may be investigated within our
approach, is the significance of intrinsic charm in the $\eta$ and 
$\eta'$ mesons. Recently a substantial charm component in the $\eta'$ 
meson has been proposed in order to explain  the large branching ratio
of the decay $B\to K\eta'$ \cite{HaZh97,ChTs97}.  Since the
$\eta\gamma$ and $\eta'\gamma$ transition form factors at large $Q^2$ 
are sensitive to intrinsic charm, our analysis  may shed further light
onto the issue of the intrinsic charm magnitude.

The paper is organized as follows: First we present the proper
expansion of pseudoscalar mesons in terms of parton Fock states and
discuss properties of the light-cone wave functions associated with
the light-quark valence Fock states (Sect.~\ref{sec2}). In Sect.\ 
\ref{sect3} we calculate the $\eta\gamma$ and $\eta'\gamma$ 
transition form factors. We employ the modified hard scattering
approach (mHSA) in which the form factors  are described by  
convolutions of perturbatively calculable
hard scattering amplitudes and non-perturbative light-cone wave
functions \cite{bot:89}. In contrast to the standard approach (sHSA)
of Brodsky and Lepage \cite{BrLe80}, the transverse momenta of the
partons and Sudakov suppressions are also taken into account in the
mHSA.  In this section we also discuss how to include the intrinsic 
charm contribution to the form factors. In Sect.~\ref{sect4} we
present numerical results on the transition form factor obtained 
on the basis of the conventional octet-singlet mixing scheme. We are 
going to demonstrate that this mixing scheme seems to be inadequate. 
A more general mixing scheme with two mixing angles is 
discussed in Sect.~\ref{sec5}. As it will turn out, this scheme
leads to a very good description of the transition form factors.
Also several other phenomenological constraints are satisfied
within this scheme. The size of an eventual contribution
from intrinsic charm is also estimated in this section.
We end the paper with our conclusions (Sect.~\ref{sec6}).
%%%%%%%%%%%%%%%%%%%%%%%%%%%%%%%%%%%%%%%%%%%%%%%%%%%%%%%%%%%%%%%%%
\vspace*{-.2cm}
\section{Fock States, Light-Cone Wave Functions and Evolution}
\label{sec2}
\setcounter{equation}{0}
\vspace*{-0.2cm}
%%%%%%%%%%%%%%%%%%%%%%%%%%%%%%%%%%%%%%%%%%%%%%%%%%%%%%%%%%%%%%%%%
For the calculation of the transition form factors within the mHSA 
we need a parton Fock state decomposition of the mesons.
Most generally, assuming isospin symmetry to be exact, we can write
($P=\eta,\eta'$)
\beq
|P\rangle\, &=&\, 
 \Psi_P^8 \,  |u\bar u + d\bar d- 2 s\bar s\rangle /\sqrt6\,
+\, \Psi_P^1 \, |u\bar u + d\bar d + s\bar s\rangle/\sqrt3\, + {}\cr&&
 \Psi_P^g \, |gg\rangle \,+\, \Psi_P^c \, |c\bar c\rangle
\,+ \ldots
\label{eq:general}
\eeq
where the light quarks are arranged in terms of the
$SU(3)_F$ octet and singlet combinations. This choice is
convenient but not mandatory. For instance, a basis where the 
$|u\bar u + d\bar d\rangle$ and $|s\bar s\rangle$ parts are treated 
separately is a reasonable choice, too. In (\ref{eq:general}) we also 
allow for gluon and charm components that may appear due to the
$U(1)_A$ anomaly. The ellipses stand for higher Fock states
with additional gluons and/or $q\bar q$ pairs.
Their contributions to the transition form factors are suppressed
by powers of $\alpha_s/Q^2$ \cite{BrLe80}, where $\alpha_s$ is the strong
coupling constant, and will therefore be neglected in our analysis.

Following \cite{JaKrRa96,JaKr93}, we write the wave functions
associated with the light-quark Fock states as ($i=8,1$)
\beq
\Psi_P^i(x,\vec k_\perp) &:=&
\frac{f_P^i}{2\sqrt6} \, \phi_P^i(x) 
\, \Sigma_P^i(\vec k_\perp/\sqrt{x\bar x}) 
\label{wfexp} \ .
\eeq
Here, the momentum fraction $x$ and the transverse momentum 
$\vec k_\perp$ refer to the quark; the antiquark momentum is 
characterized by $\bar x=1-x$ and $-\vec k_\perp$. The transverse 
momentum part, $\Sigma_P^i$, of the wave function is normalized as
\beq
\int \frac{\d^2\vec k_\perp}{16 \pi^3} \, \Sigma_P^i(\vec k_\perp/
\sqrt{x\bar x}) &=& 1 \ .
\eeq
$f_P^i$ is the decay constant of the pseudoscalar meson $P$ through 
the Fock state $i$. In the hard scattering approach, $f_P^i$ is
related to the value of the corresponding
wave function at the origin of the configuration space
\beq
\int \, \frac{\d^2\vec k_\perp}{16\pi^3} \, \int_0^1 \d x 
\, \Psi_P^i(x,\vec k_\perp)  &=&
\frac{f_P^i}{2\sqrt6}  \ .
\label{norm}
\eeq
Explicit parameterizations of the charm and gluon 
wave functions are not needed in our analysis.
The weak matrix elements that define the decay constants read
\beq
\langle0|J_{\mu5}^i|P(p)\rangle &=& \imath \, f_P^i \, p_\mu 
\label{eq:decay}
\eeq
where the axial vector currents are given by
\beq
&& J_{\mu 5}^8 =  \frac{1}{\sqrt6} \left(
  \bar u \gamma_\mu\gamma_5 u + \bar d 
  \gamma_\mu\gamma_5 d - 2\, \bar s  \gamma_\mu\gamma_5 s
  \right) \ , \quad \cr
&& J_{\mu 5}^1  =  
  \frac{1}{\sqrt3} \left(
  \bar u \gamma_\mu\gamma_5 u + \bar d 
  \gamma_\mu\gamma_5 d + \bar s  \gamma_\mu\gamma_5 s
  \right)  \ .
\eeq
The definition of the decay constants (\ref{eq:decay}) also applies to 
the charm component ($i=c$) with the current 
$J_{\mu 5}^c  = \bar c  \gamma_\mu\gamma_5 c$.

The distribution amplitudes for the octet components $\phi_P^8(x)$ 
have the same expansion upon Gegenbauer polynomials $C_n^{(3/2)}$ 
as the pion one \cite{BrLe80},
\beq
\phi_P^8(x) &=&
6 \, x \, \bar x \,
     \left\{ 1 + \sum_{n=2,4,\ldots} B_{Pn}^8(\mu_F)
       \ C_{n}^{(3/2)}(2 x - 1) \right\} \ .
\label{phiexp}
\eeq
The non-perturbative expansion coefficients evolve with
the factorization scale $\mu_F (\propto Q^2)$ as
\beq
B_{Pn}^8(\mu_F) &=&
 B_{Pn}^8(\mu_0) \,
       \left\{\frac{\alpha_s(\mu_F)}{\alpha_s(\mu_0)}\right\}^{\gamma_{n}}.
\label{b28evol} 
\eeq
Here, $\mu_0$ is a typical hadronic scale of reference for which we 
choose a value of $0.5$ GeV. Since the anomalous dimensions \
$\gamma_{n}$ are positive fractional numbers increasing with $n$, 
all distribution amplitudes evolve into
$\phi_{AS}(x) = 6 \, x \, \bar x$ asymptotically.
Rather similar forms of the octet and pion distribution
amplitudes are to be expected from symmetry considerations.
Since, to a very good approximation, the latter distribution
amplitude equals the asymptotic form\footnote
{For comparison, within the modified HSA \cite{KrRa96},
a fit of $B_{\pi2}$ to the CLEO data on the $\pi\gamma$ transition 
form factor \cite{CLEO97} yields a value of $-0.02\pm0.1$ at the
scale $\mu_0$.} this should be the case for $\phi_P^8$ too. To deal 
with eventual small deviations from the asymptotic form
it is sufficient to consider only the first non-trivial contribution 
$B_{P2}^8 \neq 0$ with $C_2^{(3/2)}(z)=3/2\,(5\, z^2 - 1)$ 
and $\gamma_2=50/81$.

In the singlet case evolution is more complicated.
The evolution equation involves an anomalous dimension
matrix which mixes the singlet and the two-gluon distribution
amplitudes. In \cite{BaGr81} the eigenfunctions and eigenvalues
of the evolution equation have been calculated.
The results for three flavors read
\beq
\phi_P^1(x) &=& 6  \, x  \bar x \,
  \left\{ 1 + {} \sum_{n=2,4,\ldots} 
         \left[B_{Pn}^{1}(\mu_F) +  \rho_n^g \, B_{Pn}^{g}(\mu_F)\right]
         \, C_n^{(3/2)}(2x-1) \right\} \nonumber \\[0.2em]
\phi_P^g(x) &=&  (x \, \bar x)^2 \,
  \sum_{n=2,4,\ldots} 
         \left[\rho_n^1 \, B_{Pn}^{1}(\mu_F) + B_{Pn}^{g}(\mu_F)\right]
         \, C_{n-1}^{(5/2)}(2x-1)\, . 
\label{singletevol}
\eeq
The indices $1$ and $g$ on the r.h.s.\ of this equation characterize
the two eigenfunctions.
A common factor $f_P^1/2\sqrt6$ is pulled out of both the distribution
amplitudes, see (\ref{wfexp}). Since only Gegenbauer polynomials 
$C_n^{(5/2)}$ of odd order contribute to the the gluon
distribution amplitude it possesses the properties
$\phi_P^g(x)= - \phi_P^g(\bar x)$ and
$\int_0^1\phi_P^g(x)\, \d x = 0$ while $\phi_P^i(x)=
\phi_P^i(\bar x)$ and $\int_0^1\phi_P^i(x)\, \d x = 1$ for
the light quarks. The interesting point is that once $\phi_P^1$ 
is determined, say from experiment, the corresponding gluon distribution 
amplitude is, in principle, fixed by evolution. This situation is 
quite similar to the one in deep inelastic lepton-hadron scattering. 
In particular, if $\phi_P^1 = \phi_{\rm AS}$ then $\phi_P^g=0$.
For the case $n=2$  we quote the numerical values of the anomalous 
dimensions $\gamma_n^{1,g}$, controlling the evolution of the singlet 
and gluon distribution amplitudes analogue to (\ref{b28evol}), and 
the coefficients $\rho_n^{1,g}$ in the eigenfunctions that are induced
by the gluon/quark admixtures ( $C_1^{(5/2)}(z)=5\, z$ ):
\beq
&& \gamma_2^1=0.59 \ ; \quad \rho_2^1 = 1.42 \ ; 
\quad
\gamma_2^g=1.24 \ ; \quad \rho_2^g =- 0.025 \ . 
\eeq
Finally, following \cite{JaKrRa96,JaKr93,BHL},
the transverse shape of the wave function is chosen to be a simple 
Gaussian ($i=1,8$) 
\beq
\Sigma_P^i(\vec k_\perp/\sqrt{x\bar x})
&=& \frac{16 \pi^2 \, (a_P^i)^2}{x\, \bar x} 
\, \exp\left[ - \frac{(a_P^i)^2\, \vec k_\perp^2}{x \, \bar x} \right]
\ . \qquad 
\label{eq:sigma}
\eeq
In the case of the pion the transverse size parameter 
$a_\pi$ is fixed through the constraint \cite{BHL}
\be
\int \d x \, \, \Psi_\pi(x,0) \,=\, \sqrt6/f_\pi \ .
\label{pirel}
\ee
That relation leads to the closed formula
$a_\pi^{-2} = (1 + B_{\pi2}(\mu_F)) \, 8 \pi^2\, f_\pi^2$
under the assumption $B_{\pi n}=0$ for $n \geq 2$.
For the asymptotic distribution amplitude one obtains  
$a_\pi = 0.86$~GeV$^{-1}$ corresponding to a r.m.s.\ transverse 
momentum of $370~{\rm MeV}$. For simplicity, we assume 
$a_P^i= a_\pi$, $i=1,8$ throughout this work. The present
data do not allow to detect differences between the
individual transverse size parameters.

We note that, leaving aside the intrinsic gluon and charm
components, the Fock state decomposition (\ref{eq:general}) 
already includes four independent wave functions that characterize the
light-quark contributions to the $\eta$ and $\eta'$ mesons, 
each in principle with its own distribution amplitude, transverse 
shape function and value  at the origin.
It is of course a formidable task to determine all the
parameters, that enter the wave functions $\Psi_P^i$, completely 
from phenomenological constraints. As already mentioned in the
Introduction one needs additional assumptions in order to simplify 
the analysis (see Sect.\ \ref{sect4}, \ref{sec5}).
%%%%%%%%%%%%%%%%%%%%%%%%%%%%%%%%%%%%%%%%%%%%%%%%%%%%%%%%%%%%%%%%%%%
\vspace*{-0.2cm}
\section{Meson-Photon Transition Form Factor}
\label{sect3}
\setcounter{equation}{0}
\vspace*{-0.2cm}
%%%%%%%%%%%%%%%%%%%%%%%%%%%%%%%%%%%%%%%%%%%%%%%%%%%%%%%%%%%%%%%%%%%
\begin{figure}[t]
\unitlength0.7cm
\begin{center}
\fmfframe(0.1,0.3)(0.1,0.5){\parbox{4cm}{\begin{fmfgraph}(5,3)
  \fmfpen{thick} \fmfleft{q1,q2} \fmfright{c,cbar}
  \fmf{photon}{q1,v1}
  \fmf{photon}{q2,v2}
  \fmf{fermion}{v1,c} 
  \fmf{fermion}{cbar,v2}
  \fmf{fermion}{v2,v1}
  \fmfdotn{v}{2}
  \end{fmfgraph}}}
\hskip2em
\fmfframe(0.1,0.3)(0.1,0.5){  \parbox{4cm}{\begin{fmfgraph}(5,3)
  \fmfpen{thick} \fmfleft{q2,q1} \fmfright{c,cbar}
   \fmf{phantom}{q2,v1} \fmf{phantom}{q1,v2}
  \fmf{photon,tension=0}{q1,v1}
  \fmf{photon,tension=0}{q2,v2}
  \fmf{fermion}{v1,c} 
  \fmf{fermion}{cbar,v2}
  \fmf{fermion}{v2,v1}
  \fmfdotn{v}{2}
  \end{fmfgraph}}}
\end{center}
\caption{Feynman graphs that determine the
leading order hard scattering amplitude.}
\label{Feynmfig}
\end{figure}
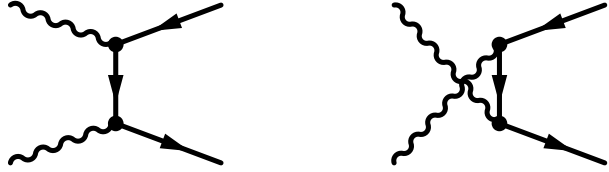

We write the $P\gamma\gamma^*$-vertex (see Fig.\ \ref{Profig}) as 
$$
\Gamma_\mu(q_1^2=0,q_2^2=-Q^2) =
i \, e_0^2 \, F_{P\gamma}(Q^2) \,
\epsilon_{\mu\nu\kappa\lambda} \, p^\nu \, q_1^\kappa \, 
\varepsilon^\lambda \ .
$$
Following \cite{KrRa96,JaKrRa96}, we
calculate the $P\gamma$ transition form factors
($P=\eta,\eta'$) for $Q^2 \geq 1$~GeV$^2$ within the
mHSA. In that approach the transverse momentum dependence
of the hard scattering amplitude is retained, and 
Sudakov suppressions are taken into account in contrast
to the sHSA.
Each quark term in (\ref{eq:general}) gives rise to an additive
contribution to the $P\gamma$ transition form factor which
is represented by a convolution of the corresponding
wave function, the hard scattering amplitude and a Sudakov
factor ($i=8,1,c$):
\beq
F_{P\gamma}^i(Q^2) &=& 
\int_0^1 \d x \int \frac{\d^2 b}{4 \pi} \, 
\hat \Psi_P^i(x,\vec b) \, \hat T_P^i(x,\vec b,Q)  \,
\exp\left[ - {\mathcal S}(x,\vec b,Q) \right]
\ ,
\label{convol}
\eeq
where $\vec b$, canonically conjugated to the transverse
momentum, is the quark-antiquark separation in the 
transverse configuration space. $\hat \Psi_P^i$ and $\hat T_P^i$ 
represent the Fourier transforms of the wave functions defined in
(\ref{eq:general}) and the hard scattering amplitudes, respectively.
To lowest order of perturbative QCD, the hard scattering
amplitudes are to be calculated from the two Feynman graphs shown 
in Fig.\ \ref{Feynmfig}. For the light-quark contributions ($i=8,1$) 
one finds 
\be
T_P^i(x,\vec k_\perp,Q) \,=
\, 2 \sqrt6 \, C_i \,
\left\{ \frac{1}{x \, Q^2 + \vec k_\perp^2 + x \, \bar x\,
    M_P^2} + ( x \leftrightarrow \bar x) \right\} 
\label{Ti}
\ee
in the transverse momentum space. The charge factors read 
$C_8 = (e_u^2+e_d^2-2e_s^2)/\sqrt6$ and
$C_1 = (e_u^2+e_d^2+e_s^2)/\sqrt3$.
Since the masses $M_P$ of the $\eta$ and $\eta'$ mesons
are rather large we allow for corresponding corrections
in the hard scattering amplitudes. Due to the symmetry of the 
distribution amplitudes under $x \leftrightarrow \bar x$
the two Feynman graphs provide identical contributions.
The Fourier transformed amplitudes are proportional
to $K_0(\sqrt{x Q^2 + x \bar x \, M_P^2} \,b)$ where
$K_0$ is the modified Bessel function of order zero.

The Sudakov factor  $\exp[-{\mathcal S}(x,b,Q)]$ takes into 
account gluonic corrections not accounted for in
the QCD evolution of the wave functions.
In the Sudakov factor $b$ plays the role of an infrared
cut-off; it sets up the interface between the non-perturbative
soft gluon contributions -- still  contained in the hadronic
wave function -- and perturbative soft gluon contributions
accounted for by the Sudakov factor.
The gliding factorization scale to be used in the evolution
of the wave functions is, hence, chosen to be $\mu_F=1/b$.
The Sudakov factor has been
calculated by Botts and Sterman~\cite{bot:89} in
next-to-leading-log approximation. The explicit
form of the Sudakov function which has been slightly improved,
can, for instance, be found in \cite{DaJaKr95}.

The charm contribution to the $P\gamma$ transition form factor
can be estimated in close analogy to the calculation of the 
$\eta_c\gamma$ form factor \cite{FeKr97a}. An important difference 
to the light-quark case consists in the mass of the charm quark 
($m_c \simeq 1.5$ GeV) that provides a second large scale in the process.
The hard scattering amplitude is therefore to be modified 
accordingly. The distribution amplitude $\phi_P^c$ is expected
to behave similar to the $\eta_c$ distribution amplitude and
should, in particular, exhibit a pronounced maximum at $x=1/2$.
It therefore suffices to use the peaking approximation 
$\phi^c = \delta(x-1/2)$. Including transverse momentum corrections 
to $O(\vec k_\perp^2/m_c^2)$, one finds for the charm contribution 
the reasonable approximation\\
\be
F_{P\gamma}^c(Q^2) \,=\,
 \frac{4 \, e_c^2 \, f_P^c}{Q^2 + M_P^2/2 + 
       2m_c^2 + 2 \langle \vec k_\perp^2\rangle_c} \ ,
\label{incharm}
\ee
where, according to \cite{FeKr97a},
a value of $710$~MeV  is used for the r.m.s.\ transverse
momentum of the charm quarks. Details of this
approximation and an assessment of its quality can be
found in \cite{FeKr97a}. Eq.\ (\ref{incharm}) possesses the highly 
welcome feature that, except of the decay constants and the r.m.s.\
transverse momentum, no further details of the charm wave function 
are required. Furthermore it represents rather an underestimate of 
the intrinsic charm contribution: Going beyond the peaking
approximation and/or inserting a value of the r.m.s.\ transverse 
momentum closer to the value of 370 MeV that we use for the light-quark
components, would even increase the magnitude of the
charm contribution.

The two-gluon components of the $\eta$ and $\eta'$ mesons
play no direct role in the analysis of the transition form factors
since their coupling to photons is suppressed by the strong
coupling constant $\alpha_s$. Moreover, the formation of a 
pseudoscalar meson from two vector particles requires orbital 
angular momentum. This implies a factor $\vec k_\perp$ in the 
corresponding spin-wave function that leads to an additional
suppression factor $\vec k_\perp^2/Q^2$ of the
two-gluon contributions. 

It is instructive to consider the asymptotic behavior of the 
transition form factors. For $\ln Q^2 \to \infty$ the Sudakov factor 
damps any contribution to the form factors except those from 
configurations with small quark-antiquark separations. Contributions 
from such configurations are actually considered in the sHSA.
Hence, the mHSA and the sHSA have the same asymptotic limit. Since,
for $\ln Q^2\to \infty$, any distribution amplitude evolves into 
the asymptotic one, the limiting behavior of the 
transition form factors (for $f_P^c=0$)
\be
Q^2 \, 
F_{P\gamma}(Q^2)  \;
\stackrel{\ln Q^2 \to \infty}{\longrightarrow} \;  
\sqrt\frac23 \, f_P^8  + \frac{4}{\sqrt3} \, f_P^1 \ .
\label{eq:asymp}
\ee
is model-independent.
\newpage  
%%%%%%%%%%%%%%%%%%%%%%%%%%%%%%%%%%%%%%%%%%%%%%%%%%%%%%%%%%
\vspace*{-0.2cm}
\section{The Octet-Singlet Mixing Scheme}
\setcounter{equation}{0}
\label{sect4}
\vspace*{-0.2cm}
%%%%%%%%%%%%%%%%%%%%%%%%%%%%%%%%%%%%%%%%%%%%%%%%%%%%%%%%%%
The most obvious way to reduce the number of parameters and 
to make contact to phenomenology, is to simplify the general ansatz 
(\ref{eq:general}) by adopting the usual  $SU(3)_F$ 
octet-singlet mixing scheme. Corresponding Fock state components of
the $\eta$ and $\eta'$ mesons are then controlled by one
and the same wave function. The octet-singlet mixing scheme is 
defined through the relations 
\beq
&& \Psi_\eta^8 = \ \ \Psi_8 \, \cos\theta_P \ , \qquad
    \Psi_{\eta'}^8 = \Psi_8 \, \sin\theta_P \ , \cr
&& \Psi_\eta^i = - \Psi_i \, \sin\theta_P \ , \qquad
   \Psi_{\eta'}^i = \Psi_i \, \cos\theta_P \ , \qquad
   (i=1,g,c)
\label{mixrel}
\eeq
for the valence Fock state wave functions defined in (\ref{eq:general}).
Eq.\ (\ref{mixrel}) implies analogous relations between the decay
constants. For the light-quark wave functions, $\Psi_8$ and $\Psi_1$, 
we use the ansatz (\ref{wfexp}). All the properties of light-quark
wave functions discussed in Sect.\ \ref{sec2} are valid for $\Psi_8$ 
and $\Psi_1$, too. The octet-singlet mixing scheme is based on the 
concept of octet $(\eta_8)$ and singlet $(\eta_1)$ mesons as $SU(3)_F$
basis states from which, by a unitary transformation, the physical 
mesons arise. This concept implies that Fock state components with 
singlet quantum numbers, in other words all Fock state components of 
the $\eta_1$ meson, contribute to the $\eta$ and $\eta'$ mesons through
the relations in the second line of (\ref{mixrel}).

Approximate $SU(3)_F$ symmetry tells us that the octet and the pion 
wave functions cannot differ much from each other. Due to the
larger quark masses involved, the octet distribution amplitude
may, at the most, be slightly more midpoint-concentrated, i.e.\ 
$B_2^8 < 0$, than the pion one which is well described by the
asymptotic form \cite{KrRa96}. The singlet wave function is not
related to the pion one by symmetry. Since the binding mechanisms of 
the quarks in the flavor octet and singlet channels are however similar,
we expect the light-quark singlet wave function $\Psi_1$
to be not too different from that of the pion. 
Thus, a reasonable starting point of the analysis of 
transition form factors is the assumption $B_{n}^i=0$, $a_i=a_\pi$
($i=1,8$; $n\geq 2$). This ansatz coincides with the one used
in \cite{JaKrRa96}. There are still three parameters to be
determined, the two decay constants $f_1$, $f_8$ and the
mixing angle. Admittedly, additional information is required
for this task since the two transition form factors do not
suffice to fix these three parameters; for any value of
the mixing angle an acceptable fit to the data can be obtained.

A useful constraint is provided by the two-photon 
decays of the $\eta$ and $\eta'$ mesons. Generalizing
the PCAC result for the $\pi^0\to\gamma\gamma$ (which
is responsible for the constraint (\ref{pirel})),
one assumes that the axial vector currents can be 
related via PCAC to the $\eta$ and $\eta'$ fields
(see e.g.\ \cite{KiPe93})
\beq
\partial^\mu J_{\mu 5}^8(z) &=&
  f_{\eta}^8 \, M_{\eta}^2 \, 
     \eta(z) + f_{\eta'}^8 \, M_{\eta'}^2 \, 
     \eta'(z) + \ldots \nonumber \\[0.2em]
\partial^\mu J_{\mu 5}^1(z) &=&
  f_{\eta}^1 \, M_{\eta}^2 \, 
     \eta(z) + f_{\eta'}^1 \, M_{\eta'}^2 \, 
     \eta'(z) + \ldots
\label{eq:currents}
\eeq
This leads to
\beq
\Gamma[\eta\phantom{'}\to\gamma\gamma] &=&
\frac{9\alpha^2}{16 \pi^3} \, M_{\eta}^3 \,
\left[\frac{C_8 \, f_{\eta'}^1 - C_1 \, f_{\eta'}^8}{
            f_{\eta'}^1 \, f_\eta^8 - f_{\eta'}^8 \, f_{\eta}^1}
\right]^2 \nonumber \ , \\[0.3em]
\Gamma[{\eta'}\to\gamma\gamma] &=&
\frac{9\alpha^2}{16 \pi^3} \, M_{{\eta'}}^3 \,
\left[\frac{-C_8 \, f_{\eta}^1 + C_1 \, f_{\eta}^8}{
            f_{\eta'}^1 \, f_\eta^8 - f_{\eta'}^8 \, f_{\eta}^1}
\right]^2 \, .
\label{eq:gammapred}
\eeq
The various decay constants appearing in (\ref{eq:gammapred}) can be 
expressed in terms of $f_1,f_8$ and $\theta_P$ by means of
(\ref{mixrel}). The experimental values of the two-photon decay widths are
\cite{PDG96}
\be
\Gamma[\eta \to \gamma\gamma] = (0.51\pm 0.026)\,{\rm keV} \, ,\:\:
   \Gamma[\eta' \to \gamma\gamma] = (4.26 \pm 0.19)\,{\rm keV}\, .\:\:
\label{2gamma}
\ee
We do not include the value $0.324\pm 0.046$ keV obtained from the
Primakoff production measurement for $\eta \to \gamma\gamma$.

\begin{figure}[t]
{\psfig{file=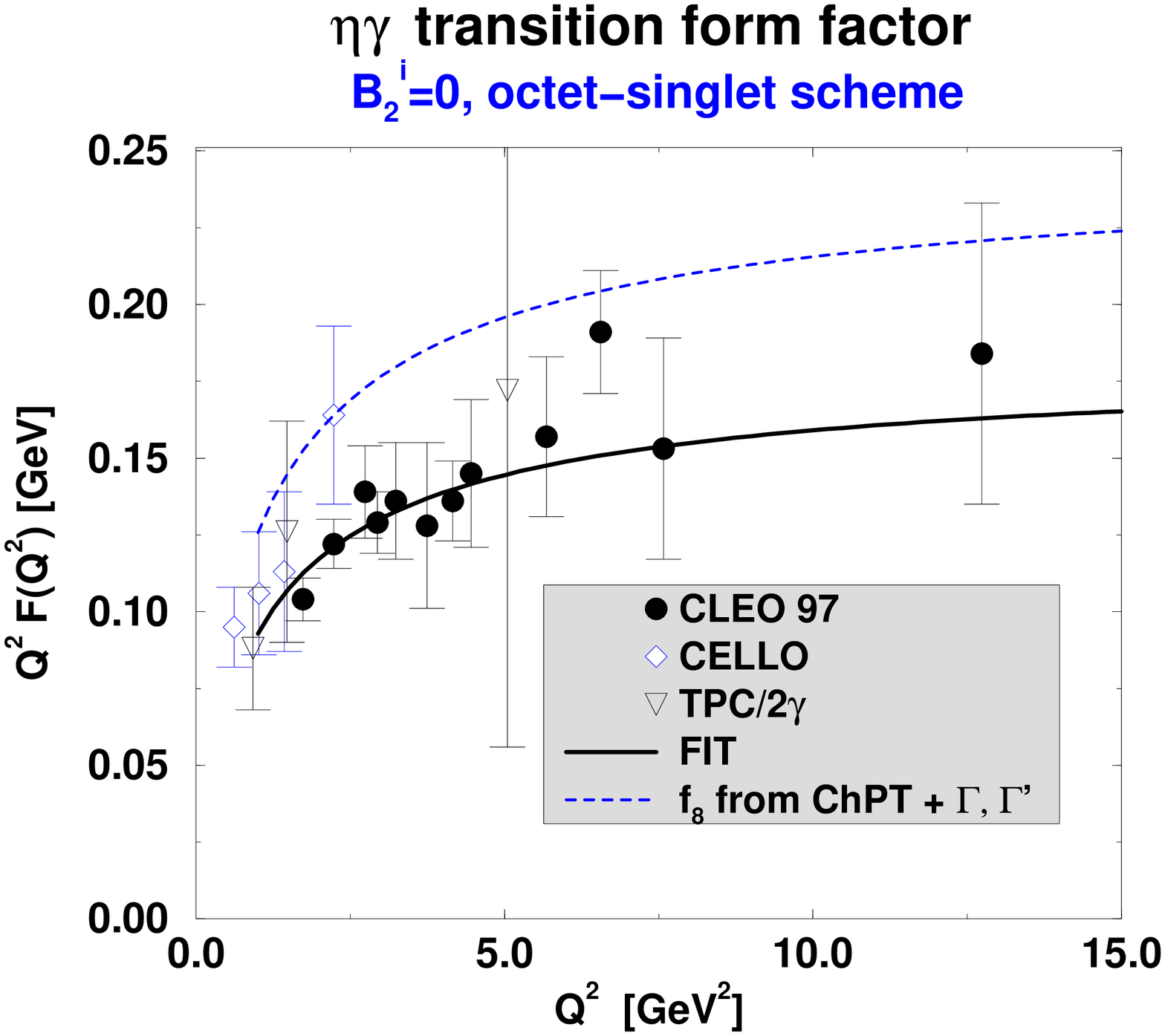, width=5.8cm,angle = -90}} \
\hskip-.5cm
{\psfig{file=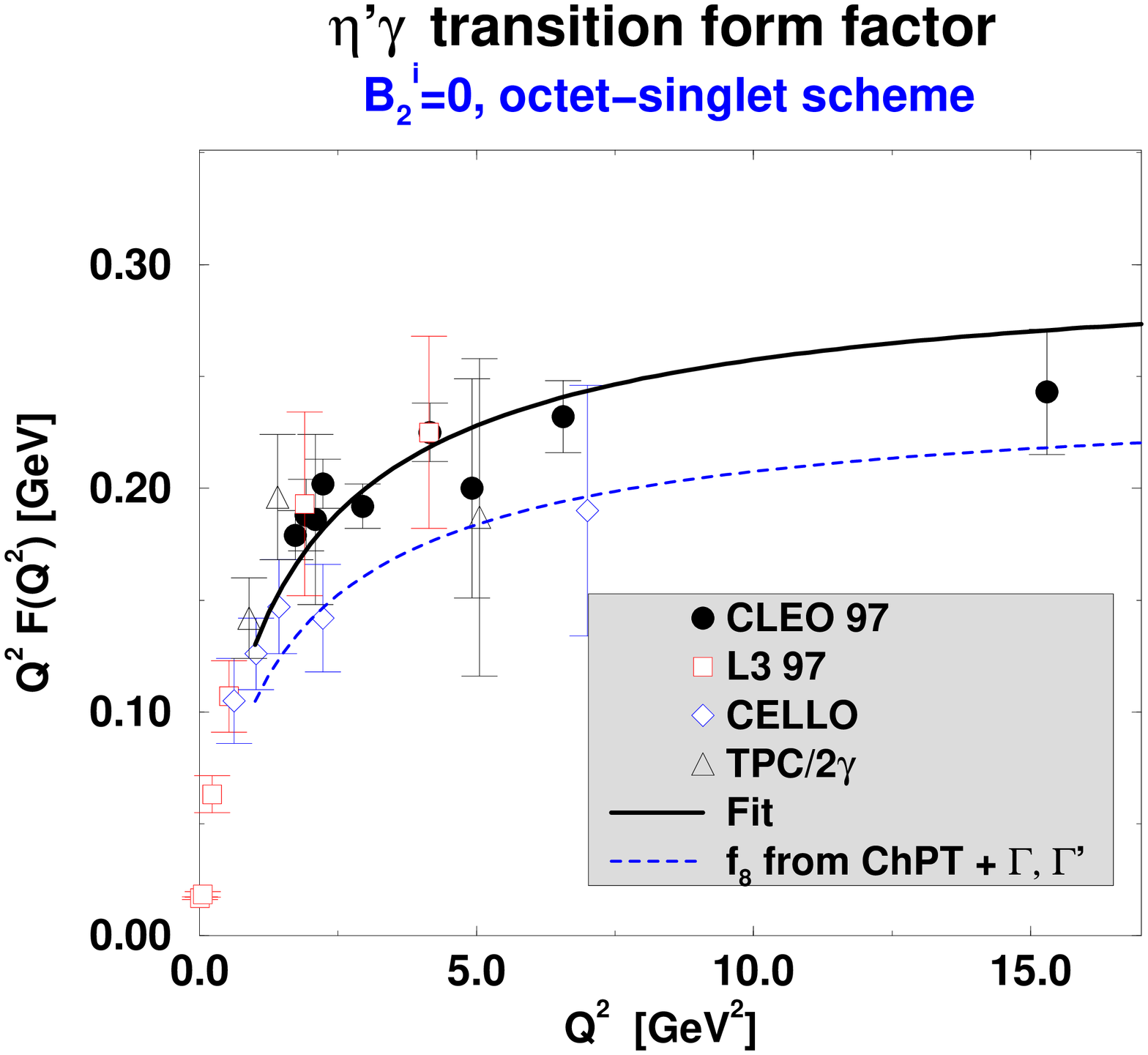, width=5.8cm,angle = -90}} 
\caption{Results for the $\eta\gamma$ and
$\eta'\gamma$ transition form factors 
in the octet-singlet scheme
($a_i=a_\pi$, $f_P^c=0$).
For comparison we also show results obtained from the
parameter set OSS. Data are taken from 
\cite{CLEO97,L397,CELLO91,TPC90}.}
\label{fig1}
\end{figure}

Using the asymptotic distribution amplitudes as well as universal 
transverse size parameters and ignoring an eventual charm
contribution, we fit the decay constants $f_1$ and $f_8$ and the mixing
angle to the transition form factor data above $1$ GeV$^2$ 
\cite{CLEO97,L397,CELLO91,TPC90} and the two-photon decay widths. The 
results of this excellent fit are shown in Tab.\ \ref{table1} and
Fig.\ \ref{fig1}. 
\begin{table}[t]
\caption{Results of the $\chi^2$ fit to the $\eta\gamma$ and
$\eta'\gamma$ transition form factors and the two-photon
decay widths in the octet-singlet
mixing scheme ($a_i=a_\pi$).
For comparison we also show results obtained from the
parameter set OSS (see text).}
\label{table1}
\hspace{0.5em}
\begin{center}
\begin{tabular}{|l|ccc||c||cc||c|}
\hline
%%%%%%%%%%%%%%%%%%%%% as_81  %%%%%%%%%%%%%%%%%%%%%%%%%%%%%%%%%%%%%%%
\multicolumn{8}{|c|}{octet-singlet scheme, $B_2^i=0$, $f_P^c=0$} \\
\hline\hline   
& $ \theta_P$ &$f_{8}/f_\pi$ & $f_1/f_\pi$ &$\chi^2/$dof &
$\Gamma_{\eta\to\gamma\gamma}$ &$\Gamma_{\eta'\to\gamma\gamma}$ &
$R_{J/\psi}$ 
\\
\hline\hline
FIT & $-15.1^\circ $ & 0.91 & 1.14 &
 26/33 &  0.50  & 4.10 & 11.2
\\
\hline
OSS &
\underline{$-22.2^\circ $} & \underline{1.28} & \underline{1.07} &
 238/34 & 0.51 & 4.26 & 4.9 
\\
\hline\hline
 &&& & &[keV] & [keV] &  \\
\hline
\end{tabular}
\end{center}
\end{table}
%\vspace{1em} 
It is interesting to compare our values for $f_1$,
$f_8$ and $\theta_P$ with those obtained from chiral perturbation
theory (ChPT) \cite{GaLe85,DoHoLi86}
\be
\theta_P \simeq - 20^\circ \ ;
\quad 
f_8 = 1.28 \, f_\pi \ ;
\quad
f_1 \simeq 1.1 \, f_\pi \ . 
\label{chptset}
\ee
Whereas $f_8$ is theoretically on sound grounds by its
relation to the pion and kaon decay constants
\be
f_8\,=\, \sqrt{\frac43 \, f_K^2 - \frac13 \, f_\pi^2} =
1.28 \, f_\pi \ ,
\label{f8rel}
\ee
the other two parameters are subject to rather large
phenomenological and theoretical uncertainties. For instance,
$f_1$ may acquire scale dependent corrections of order $1/N_c$ due to the
gluon anomaly \cite{Leutwyler97}. Our fitted set of parameters, which 
is rather similar to that one quoted in \cite{JaKrRa96}, differ from 
(\ref{chptset}) in the values of the mixing angle and the octet decay
constant. The latter discrepancy is rather serious since, as we said, 
$f_8$ is well determined from ChPT (\ref{f8rel}). In phenomenological 
analyses based on the octet-singlet mixing scheme frequently (e.g.\ 
\cite{JaKrRa96,Bramon:1997mf,BaFrTy95}), but not always (e.g.\ 
\cite{Gilman:1987ax}), values for the mixing angle are obtained that 
are smaller in modulus than the ChPT value.

There is another phenomenological test of our parameters.
According to \cite{Novikov:1980uy}, the radiative
$J/\psi\to P\gamma$ decays are dominated by  
non-perturbative gluonic matrix elements:
\be
R_{J/\psi} \,=\,  \frac{\Gamma[J/\psi \to \eta'\gamma]}
                   {\Gamma[J/\psi \to \eta\gamma]}
\, \simeq\,
\left| \frac{\langle 0|G \tilde G|\eta'\rangle}{
             \langle 0|G \tilde G|\eta \rangle}\right|^2 \,
\left(\frac{k_{\eta'}}{k_\eta}\right)^3
\label{rjpsi}
\ee
where $k_P=M_{J/\psi} \, (1- M_P^2/M_{J/\psi}^2)/2$
being the three-momentum of the $P$ meson in the
rest frame of the decaying $J/\psi$ meson (with mass $M_{J/\psi}$). 
$G$ is the gluonic field-strength tensor and $\tilde G$ its dual.
We stress that these gluonic matrix elements are not related to the 
two-gluon components of the $\eta$ and $\eta'$ mesons  
appearing in (\ref{eq:general}) and (\ref{mixrel}).
The light-quark contributions to these decays, while responsible
for the $J/\psi \to \pi\gamma$ decay, are negligible small.
Since these gluonic contributions have singlet quantum numbers, 
(\ref{rjpsi}) reduces to
\be
R_{J/\psi} 
\,=\, \cot^2\theta_P \,\left(\frac{k_{\eta'}}{k_\eta}\right)^3
\label{naive}
\ee
within the octet-singlet mixing scheme. Quite generally, the relation 
(\ref{naive}) may be regarded as an immediate consequence of singlet 
dominance. As inspection of Tab.\ \ref{table1} reveals,
a value of $-15^\circ$ for the mixing angle is in conflict with the 
experimental value of $5.0\pm0.8$ \cite{PDG96} for $R_{J/\psi}$.

The conflict between ChPT and the form factor analysis may be
further elucidated by the following test: Keeping $f_8$ fixed
at the ChPT value (\ref{f8rel}), we can determine $\theta_P$ and
$f_1$ from the two-photon widths (\ref{eq:gammapred}). The resulting
set of values, termed OSS, is listed in Tab.\ \ref{table1}. 
It is qualitatively and quantitatively equivalent to
the parameter set (\ref{chptset}). Using the set OSS,
we evaluate the transition form factors again and arrive at
very bad results (see Tab.~\ref{table1}
and Fig.~\ref{fig1}). The ratio $R_{J/\psi}$, on the other hand,
acquires a reasonable value.

One may hold the use of the asymptotic distribution amplitudes
responsible for the apparent discrepancy between ChPT
and the form factor analysis. In order to investigate this
possibility we take the OSS set of parameters,
keep $a_i=a_\pi$ as before and fit the two expansion
coefficients $B_2^i$ ($i=8,1$) to the transition form factor
data. We find $B_2^8(\mu_0)=-0.86$ and $B_2^1(\mu_0)=0.13$.
Thus, the demand of the ChPT values (\ref{chptset} or OSS) for the 
decay constants and the mixing angle still leads to a good fit to the 
transition form factors, the two-photon decay widths and $R_{J/\psi}$,
but at the expense of an octet distribution amplitude which is very
different from the asymptotic one. Such a strong modification
of $\phi_8(x)$ seems unlikely, considering the quality of
$SU(3)_F$ symmetry. It is also at variance with the recent estimate
of $B_2^8\, (\simeq -0.04)$ obtained in the analysis of
$\chi_{cJ} \to \eta\eta$ decays ($J=0,2$) \cite{Bolz:1997}.
We do not vary the values of the transverse size parameters
since their influence on the transition form factor is rather small.
They merely influence the curvature of the transition form factors
at smaller values of momentum transfer.

From these considerations we conclude that the
octet-singlet scheme for the $\eta$-$\eta'$
system, although quite attractive due to
its simplicity in phenomenological analyses, seems to
be inadequate and should perhaps be given up in
favor of a more general description of the $\eta$-$\eta'$
system. We are going to investigate such a 
scheme in the next section.
%%%%%%%%%%%%%%%%%%%%%%%%%%%%%%%%%%%%%%%%%%%%%%%%%%%%%%%%%%%%%%%
\vspace*{-0.2cm}
\section{The Two-Angle Mixing Scheme}
\setcounter{equation}{0}
\label{sec5}
\vspace*{-0.2cm}
%%%%%%%%%%%%%%%%%%%%%%%%%%%%%%%%%%%%%%%%%%%%%%%%%%%%%%%%%%%%%%%
We define this new mixing scheme through the relations
\beq
&& \Psi_\eta^8 = \ \ \Psi_8 \, \cos\theta_8 \ , \qquad
    \Psi_{\eta'}^8 = \Psi_8 \, \sin\theta_8 \ , \cr
&& \Psi_\eta^i = - \Psi_i \, \sin\theta_1 \ , \qquad
   \Psi_{\eta'}^i = \Psi_i \, \cos\theta_1 \ , \qquad
   (i=1,g,c) \ .
\label{newmixrel}
\eeq
The analogous relations of the
light-quark decay constants $f_P^i$ have been introduced by Leutwyler 
\cite{Leutwyler97} and, in a somewhat different parametrization, by
Kiselev and Petrov \cite{KiPe93}. 

\begin{figure}[t]
{\psfig{file=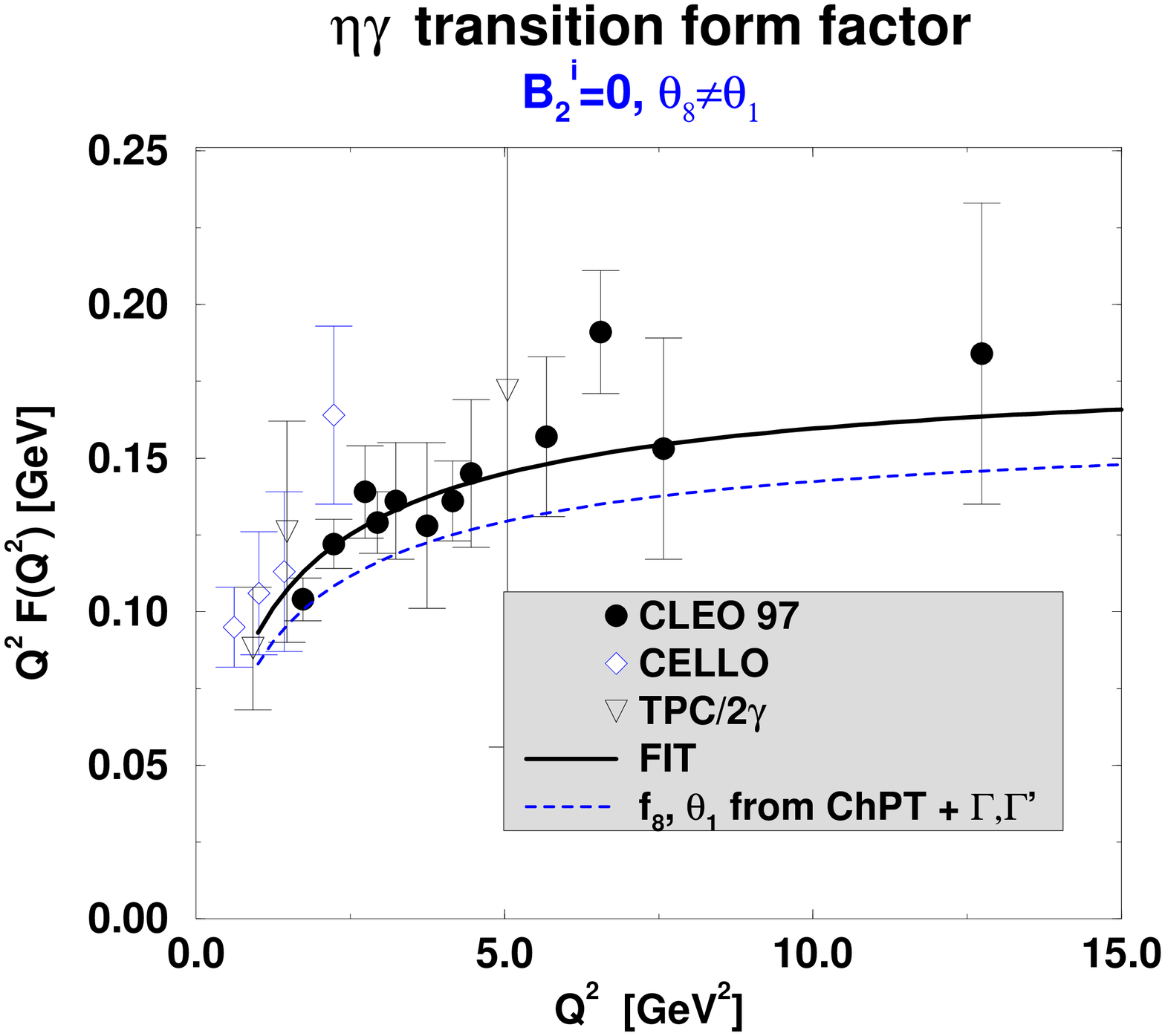, width=5.8cm,angle = -90}} \
\hskip-.5cm
{\psfig{file=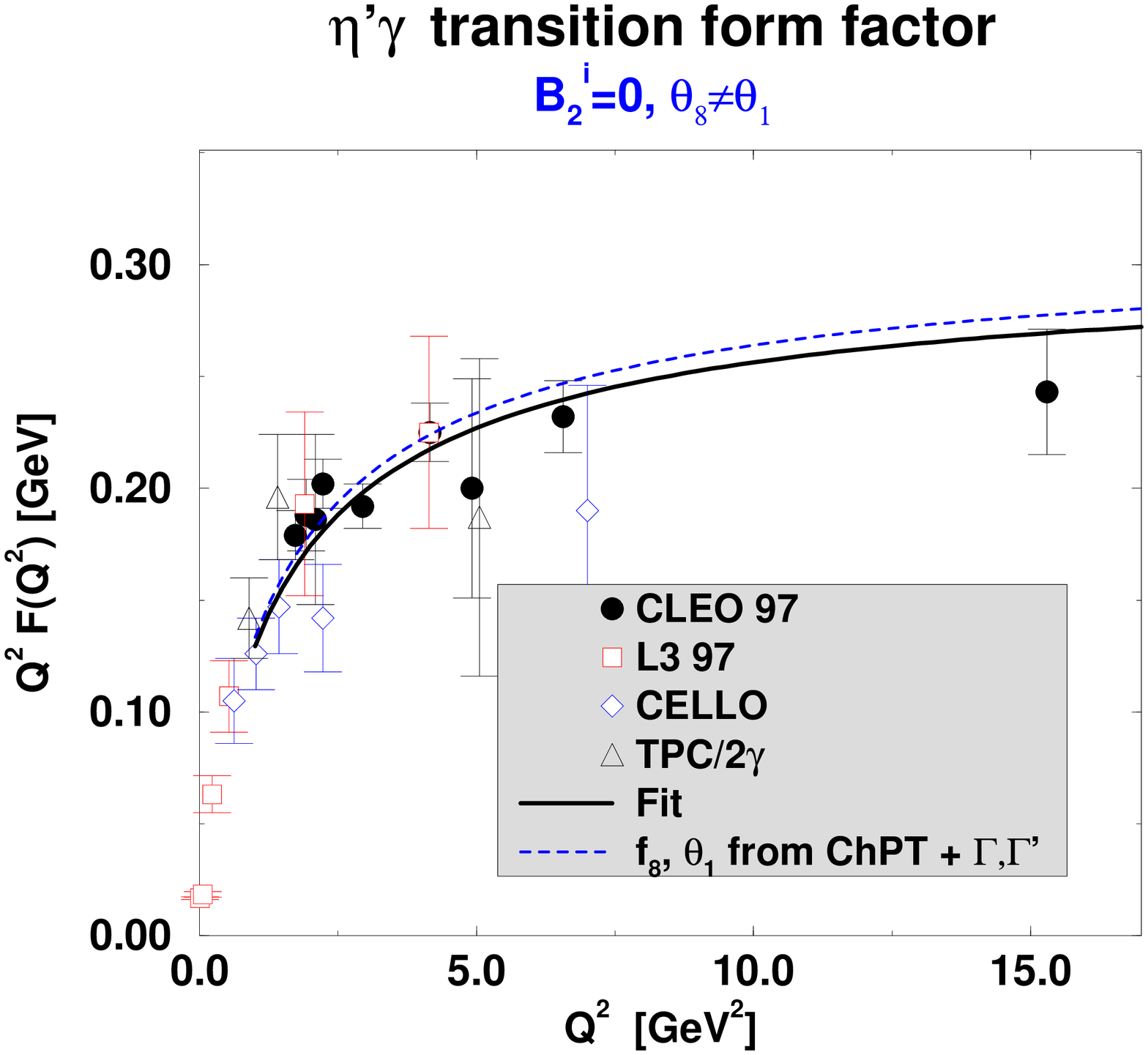, width=5.8cm,angle = -90}} 
\caption{Results for the $\eta\gamma$ and
$\eta'\gamma$ transition form factors in the two-angle mixing scheme
($a_i=a_\pi$, $f_P^c=0$). For comparison we also show results obtained from
the parameter set TAS. Data
are taken from \cite{CLEO97,L397,CELLO91,TPC90}. }
\label{fig1a}
\end{figure}

Again, we are using the asymptotic distribution amplitudes and the 
universal transverse size parameters in the analysis of the transition
form factors. Since we now have to determine one more parameter 
we need one more constraint. Thus, 
besides the two-photon decay widths (\ref{eq:gammapred}), 
we use the theoretically reliable ChPT relation (\ref{f8rel}), i.e.\ 
we take $f_8=1.28 \, f_\pi$. The results of that fit are shown
in Tab.~\ref{table1a} and Fig.~\ref{fig1a}. As for the
\begin{table}[t]
\caption{Results of the $\chi^2$ fit to the $\eta\gamma$ and
$\eta'\gamma$ transition form factors and the two-photon widths
in the two-angle mixing scheme ($a_i=a_\pi$). 
Underlined parameters are kept fixed in the fit.
For comparison we also show results obtained from
the parameter set TAS (see text).}
\label{table1a}
\hspace{0.5em}
\begin{center}
\begin{tabular}{|l|cccc||c||cc||c|}
\hline
%%%%%%%%%%%%%%%%%%%%% as_81  %%%%%%%%%%%%%%%%%%%%%%%%%%%%%%%%%%%%%%%
\multicolumn{9}{|c|}{$\theta_8\neq \theta_1$, $B_2^i=0$, $f_P^c=0$} \\
\hline\hline   
& $ \theta_8$ & $\theta_1$ & $f_{8}/f_\pi$ & $f_1/f_\pi$ &$\chi^2/$dof &
$\Gamma_{\eta\to\gamma\gamma}$ &$\Gamma_{\eta'\to\gamma\gamma}$ &
$R_{J/\psi}$
\\
\hline\hline
FIT & $-22.2^\circ $ & $-9.1^\circ $ & \underline{1.28} & 1.20 &
 26/33 &  0.50  & 4.11 & 5.1
\\
\hline
TAS &
\underline{$-22.2^\circ $} &\underline{$-5.9^\circ $} &
 \underline{1.28} & \underline{1.22} &
 41/34 &  0.51 & 4.26 & 6.2 
\\
\hline\hline
 &&&& & &[keV] & [keV] &  \\
\hline
\end{tabular}
\end{center}
\end{table}
%\vspace{2em} 
octet-singlet mixing scheme the results are very good and
agree now quite well with a recent determination of these
parameters from ChPT within the new mixing scheme \cite{Leutwyler97}:
\beq
&&
\theta_8 = -20.5^\circ , \ \theta_1 \simeq -4^\circ , \
f_8=1.28 \, f_\pi , \
f_1 \simeq 1.25 \, f_\pi .
\label{leutset}
\eeq
The only noticeable deviation between (\ref{leutset}) and
the fitted parameters is to be observed for the angle $\theta_1$
which, within ChPT, is related to the $SU(3)_F$ breaking effects
in the decay constants of the pseudoscalar octet \cite{Leutwyler97}
\be
\sin(\theta_1-\theta_8) \;\simeq\;
\frac{2\sqrt2 \, (f_K^2-f_\pi^2)}{3 \, f_8^2} \ .
\label{anglediff}
\ee
This approximately valid relation leads to a difference
$\theta_1-\theta_8$ of the mixing angles of about
$16^\circ$, larger but, in regard to the uncertainties in
(\ref{anglediff}), not in conflict with the the fitted value
of $13^\circ$.

The ratio of the $J/\psi\to P\gamma$ decay widths can still be
cast into the form (\ref{naive}), but the angle appearing
there is now to be understood as the mixing angle of the
non-perturbative gluon contribution which may -- and should -- 
differ from the angle $\theta_1$ defined in (\ref{newmixrel}).
In \cite{KiPe93,BaFrTy95} this angle, and hence $R_{J/\psi}$,
has been estimated using PCAC and taking into account a
substantial non-zero strange quark mass (whereas $m_u,m_d \simeq 0$).
This is well in the spirit of the new mixing scheme (\ref{newmixrel}) where
large $SU(3)_F$ breaking effects induce the substantial difference between 
the two mixing angles $\theta_8$ and $\theta_1$ (\ref{anglediff}).
In our notation the result of \cite{KiPe93} reads
\be
R_{J/\psi}\,=\,
\left|\frac{M_{\eta'}^2 \, (f_{\eta'}^8  + \sqrt2 \, f_{\eta'}^1)}{
      M_\eta^2    \, (f_\eta^8     + \sqrt2 \, f_\eta^1)}\right|^2\,
\left(\frac{k_{\eta'}}{k_\eta}\right)^3
\label{balletal}
\ee
Inserting the fitted values of the parameters quoted in
Tab.~\ref{table1a} into (\ref{balletal}), one obtains a value
of $5.1$ for $R_{J/\psi}$ in perfect agreement with experiment.
For comparison, we also show results in Tab.~\ref{table1a} and in 
Fig.~\ref{fig1a} that are evaluated with a set of parameters 
determined from the two ChPT relations (\ref{f8rel}) and 
(\ref{anglediff}) as well as from the two-photon decay widths 
(\ref{eq:gammapred}).
This set of parameters, termed TAS, is, not surprisingly,
close to the fitted values as well as to the ChPT values 
(\ref{leutset}). It leads to a somewhat worse fit but is not in 
severe disagreement with the transition form factor data\footnote{The
set of parameters quoted in \cite{KiPe93} differs from the sets 
(\ref{leutset}), TAS and the fitted one (see Tab.\ 
\ref{table1a}) substantially and is not consistent with
the $\eta\gamma$ and $\eta'\gamma$ transition form factors.}. 
Also the value of $R_{J/\psi}$ is only about one standard deviation
above the experimental result. The agreement with the transition form 
factor data can be improved in this case by allowing for
non-zero Gegenbauer coefficients. Values of $B_2^8(\mu_0)\simeq 0.25$
and $B_2^1(\mu_0) \simeq 0$ lead to a reasonable fit of the data.
\par
Recently, substantial intrinsic charm in the $\eta'$ meson
has been proposed \cite{HaZh97,ChTs97} in order to explain the large
branching ratios of the decays $B\to K\eta'$ and $B\to X_s\eta'$.
From the experimental measurements the authors of \cite{HaZh97} obtain 
an absolute value of $140$~MeV for the charm decay constant 
$f_{\eta'}^c$ (see (\ref{eq:decay})). This surprisingly large value is
claimed to be justified within a QCD sum rule analysis.
Another analysis of $B$ decays \cite{ChTs97} yields the more moderate
value of $-50$~MeV for $f_{\eta'}^c$. If $f_{\eta'}^c$ is that large, 
the radiative decay $J/\psi \to \eta'\gamma$ may be dominated
by a contribution where the $c\bar{c}$ pair runs from the $J/\psi$ to
the $\eta'$ meson instead of being annihilated. On that supposition 
the width of that process can be calculated along the same
lines as that one for the $J/\psi\to\eta_c\gamma$ decay. The
ratio of the two decay widths reads
\be
\frac{\Gamma[J/\psi \to \eta'\gamma]}
     {\Gamma[J/\psi \to \eta_c\gamma]} \,=\,
\kappa^2 \, \left(\frac{f_{\eta'}^c}{f_{\eta_c}}\right)^2 \,
\left(\frac{k_{\eta'}}{k_{\eta_c}}\right)^3
\label{AlGr}
\ee
in analogy to (\ref{naive}). $\kappa$ represents the ratio of
the $J/\psi$-$\eta'$ and $J/\psi$-$\eta_c$ wave function
overlaps. In \cite{AlGr97} $\kappa$ was assumed to be unity,
i.e.\ both the $\eta'$ and the $\eta_c$ mesons basically
behave like non-relativistic bound states of heavy quarks
with about the same overlap with the $J/\psi$ wave function
(each close to unity).
From the experimental values of the decay widths one then
estimates $|f_{\eta'}^c|= 6$~MeV in contradiction to the
initial assumption. In regard to the large binding energy
required for the $c\bar c$ component of the $\eta'$ meson,
a value of $\kappa$ significantly less than unity seems not
implausible. Consequently, a value larger than 6~MeV for
$f_{\eta'}^c$ cannot really be excluded by means of (\ref{AlGr}).

\begin{figure}[t]
{\psfig{file=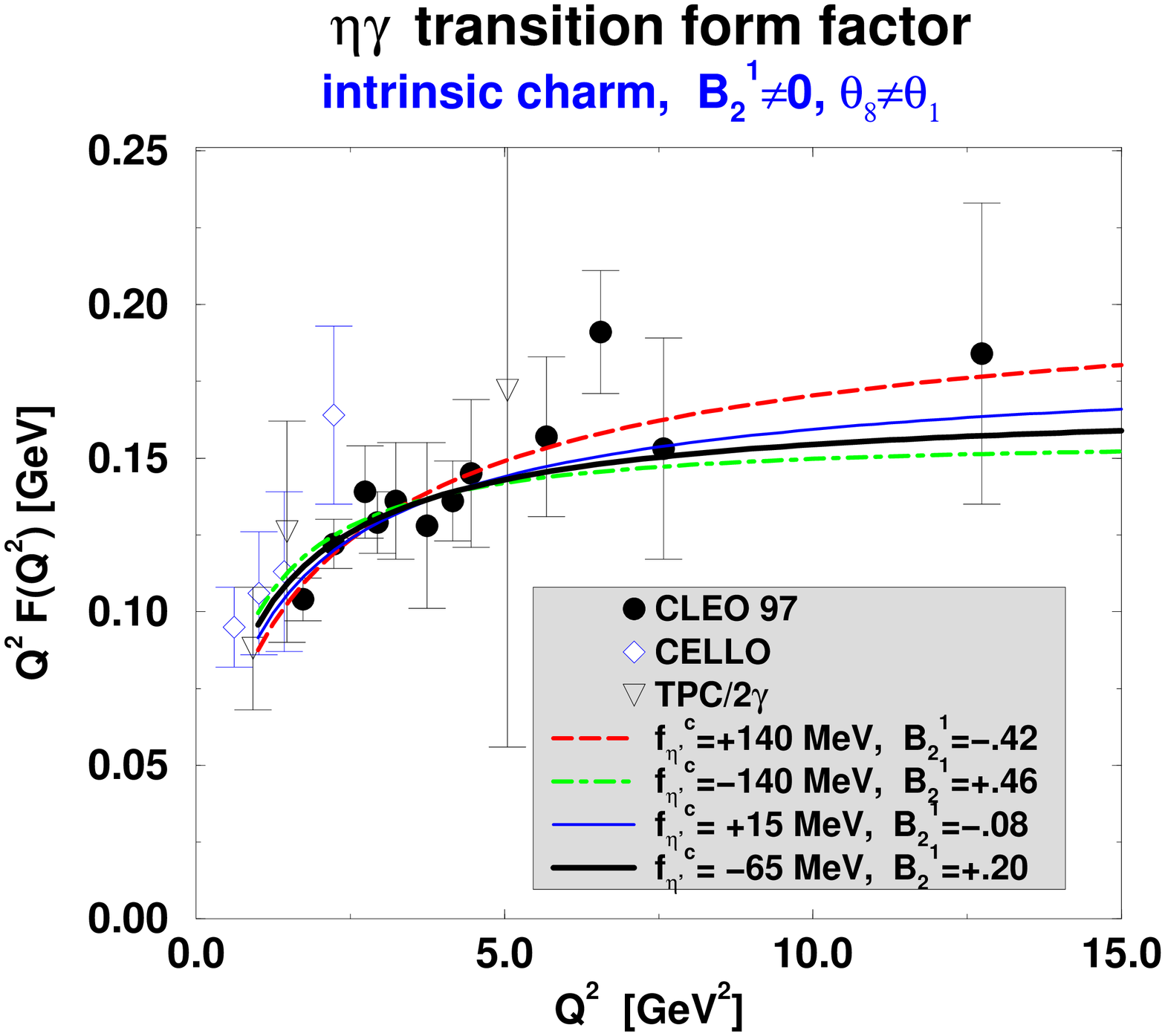, width=5.8cm,angle = -90}} \
\hskip-.5cm
{\psfig{file=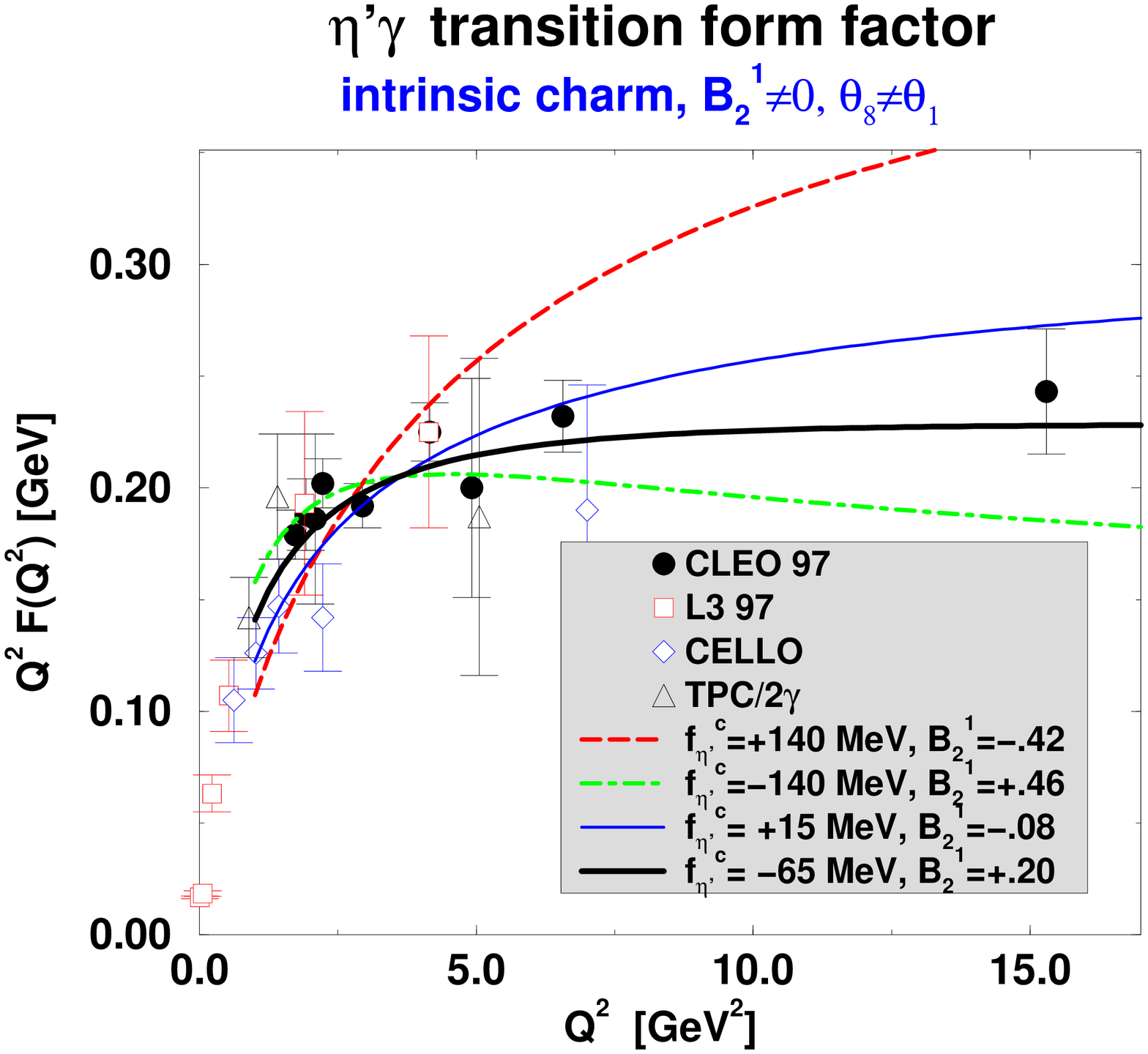, width=5.8cm,angle = -90}} 
\caption{Sample results for the $\eta\gamma$
and $\eta'\gamma$ transition form factors 
obtained from light quark
(with the parameter set FIT quoted in Tab.~\ref{table1a},
$B_2^8=0$ and $a_i=a_\pi$) and charm contributions. Data are taken from 
\cite{CLEO97,L397,CELLO91,TPC90}.}
\label{figcc}
\end{figure}

We are now going to estimate the size of the intrinsic charm.
Since the effect of the charm component of the $\eta$ meson
is suppressed by the small singlet mixing angle ($f_{\eta}^c
= - \tan\theta_1 \, f_{\eta'}^c$, see (\ref{newmixrel})) 
we can mainly concentrate ourselves 
on the $\eta'\gamma$ transition form factor in the
following discussion. The large charm quark mass effectuates
a strong suppression of the charm contribution to the 
$\eta'\gamma$ form factor at small values of $Q^2$
(see (\ref{incharm})); the main effect of it shows up for,
say, $Q^2 \gsim 4$~GeV$^2$. The charm contribution
therefore approaches its asymptotic behavior with a much
slower rate than the light-quark contributions.
This difference in the curvature is the crucial point that
allows to disentangle the charm and the light-quark
contributions. In order to determine the range of allowed
$f_{\eta'}^c$ values, we fit this parameter as well as
$B_2^1$ to the $\eta'\gamma$ and $\eta\gamma$ form factors, keeping the 
parameter set FIT given in Tab.~\ref{table1a} fixed. $f_{\eta}^c$ is
determined by means of (\ref{newmixrel}). The variation of $B_2^1$ 
changes the strength of the singlet part of the light-quark 
contributions, that is dominant in the, for these considerations,
most important $\eta'$ case, and thus makes space for a charm 
contribution. We could have freed $f_1$ instead of $B_2^1$, but this 
procedure would have the disadvantage of eventually
destroying the agreement of our results with ChPT and the
good description of the two-photon decay widths.
The numerical analysis yields the following range of
allowed values for $f_{\eta'}^c$
\be
  -65\,{\rm MeV} \leq f_{\eta'}^c \leq 15\, {\rm MeV}
\label{ccbound}
\ee
The corresponding changes of $B_2^1$ are moderate (see
Fig.~\ref{figcc})
%\vspace{2em} 
and do not lead to implausible singlet distribution amplitudes.
The results for the form factors obtained with the values
$15$ and $-65$~MeV for $f_{\eta'}^c$ are shown in Fig.~\ref{figcc}.
For comparison,
results with $\pm 140$~MeV are also shown in this figure. 
The latter two values, which require drastic changes of $B_2^1$,
lead to results for the $\eta'\gamma$ transition form factor
in clear conflict with the data above $4$~GeV$^2$.
More restrictive bounds on $f_{\eta'}^c$ than (\ref{ccbound})
require more form factor data above $4$~GeV$^2$.
We stress that the approximation (\ref{incharm}) rather underestimates
the effect of intrinsic charm (see Sect.~\ref{sect3}).
Finally, we comment on $R_{J/\psi}$ and emphasize that this
quantity is not included in our fits; it is only used as an
accompanying test of the results. With the large range
(\ref{ccbound}) of allowed $f_{\eta'}^c$ values the gluon
dominance, assumed up to now for the $J/\psi\to\eta\gamma$
and $J/\psi\to\eta'\gamma$ decays,
may not be true anymore, and, hence, the mentioned
successful test is perhaps accidental.
Whether the $J/\psi\to \eta(\eta')\gamma$ decay can reliably  be
explained by intrinsic charm in the
$\eta(\eta')$ meson remains to be shown.
%%%%%%%%%%%%%%%%%%%%%%%%%%%%%%%%%%%%%%%%%%%%%%%%%%%%%%%%%%%%%%%%%%%%%%%%
\vspace*{-0.2cm}
\section{Conclusions}
\label{sec6}
\setcounter{equation}{0}
\vspace*{-0.2cm}
%%%%%%%%%%%%%%%%%%%%%%%%%%%%%%%%%%%%%%%%%%%%%%%%%%%%%%%%%%%%%%%%%%%%%%%%
The modified hard scattering approach is shown
to provide a consistent description of the $\eta\gamma$
and $\eta'\gamma$ transition form factors over a wide
range of the momentum transfer $1$~GeV$^2\leq Q^2\leq 15$~GeV$^2$,
where experimental data is now available. 
It is to be emphasized that this result is not trivial at all:
The rather strong deviations from the dimensional
counting behavior ($Q^2 \, F_{P\gamma} \simeq {\rm const.}$)
at small momentum transfer appear as a consquence of the 
transverse momentum dependence and the Sudakov suppressions
included in the mHSA. We have included the two-photon decay widths 
of the $\eta$ and  $\eta'$ mesons in the analysis in order to
reduce the degrees of freedom in the analysis. The ratio $R_{J/\psi}$
of the $J\psi \to \eta'\gamma$ and $J/\psi \to \eta\gamma$
decay widths is merely used as an additional test.

Our analysis of the transition form factors
allows us to extract interesting information
on the $\eta$ and $\eta'$ wave functions.
Our starting point is the assumption that, like the pion,
the light-quark components of the $\eta$ and $\eta'$
mesons are described by the asymptotic form of the
distribution amplitude and a Gaussian transverse momentum
dependence with a universal transverse size parameter,
$a_i=a_\pi$ ($i=1,8$). The wave functions at the origin
of the configuration space, or the corresponding decay 
constants, are considered as free parameters to be
determined from the analysis. We found hints at an inadequacy of
the conventional octet-singlet scheme used to describe the
mixing and the $SU(3)_F$ symmetry breaking in the $\eta$-$\eta'$
system. The fit yields values of the mixing angle and the
octet decay constant which substantially deviate from the
ChPT values. Moreover, the fitted set of parameters,
quoted in Tab.~\ref{table1}, does not pass the $R_{J/\psi}$
test. Agreement with ChPT can only be obtained at the expense
of large deviations of the octet wave function from the
pion one. With regard to the quality of $SU(3)_F$
symmetry this seems to be unrealistic.

In contrast to the conventional octet-singlet mixing scheme,
the more general two-angle mixing scheme \cite{Leutwyler97,KiPe93} 
meets all requirements: It provides a very good description of the
transition form factors with the asymptotic form of the
distribution amplitudes. The set of parameters
\beq
&&
\theta_8 = -22.2^\circ , \quad
\theta_1 \simeq -9.1^\circ , \quad
f_8=1.28 \, f_\pi , \quad 
f_1 \simeq 1.20 \, f_\pi 
\label{finalset}
\eeq
is in reasonable agreement with the recent ChPT results
obtained within that new mixing scheme \cite{Leutwyler97} and
reproduces the two-photon decay widths of the $\eta$
and $\eta'$ mesons as well as the ratio $R_{J/\psi}$.
The parameter set (\ref{finalset}) implies the
asymptotic behavior of the transition 
form factors $Q^2 \, F_{P\gamma} \to 184$~MeV and $306 $~MeV for
the $\eta$ and $\eta'$ cases, respectively. Thus,
numerically the $\eta\gamma$ and $\pi\gamma$
transition form factor have the same asymptotic values.

Finally, we have also investigated whether a large intrinsic
charm component in the $\eta$ and $\eta'$ mesons is allowed by the 
transition form factor data. From our analysis we estimate the range
of allowed $f_{\eta'}^c$ values to be: $-65 {\rm MeV} \leq
f_{\eta'}^c \leq 15$ MeV. Values as large as $140$ MeV,
as suggested in \cite{HaZh97}, seem to be excluded.
More restrictive bounds on $f_{\eta'}^c$ require more
transition form factor data above $4$~GeV$^2$.
%%%%%%%%%%%%%%%%%%%%%%%%%%%%%%%%%%%%%%%%%%%%%%%%%%%%%%%%%%%%%%%%%
\vspace*{-0.2cm}
\section*{Acknowledgments}
\vspace*{-0.2cm}
%%%%%%%%%%%%%%%%%%%%%%%%%%%%%%%%%%%%%%%%%%%%%%%%%%%%%%%%%%%%%%%%%
We would like to thank V. Savinov and M. Kienzle 
for providing us with the recent experimental data.
One of us (T.F.) is also grateful to H. Leutwyler and
A. Ali for fruitful discussions.
This work is supported in part by the European TMR network
contract {\sc ERB-4061-PI95}.

\end{fmffile}
\end{document}